\documentclass[11pt]{report}

\usepackage{amsmath,amssymb,amsfonts}
\usepackage{mathrsfs}
\usepackage{bm}
\usepackage{tabularx}
\usepackage{graphicx}
\usepackage[dvipsnames]{xcolor}
\usepackage{hyperref}
\hypersetup{colorlinks=true,citecolor=[RGB]{22 36 62},linkcolor=[RGB]{22 36 62},urlcolor=[RGB]{22 36 62}}
\usepackage[T1]{fontenc}
\usepackage{textcomp}

\RequirePackage[procnames]{listings}
\definecolor{keywordclr}{rgb}{0, 0.2, 0.667}      
\definecolor{fnctionclr}{rgb}{0.467, 0, 0.533}
\definecolor{builtinclr}{rgb}{0.467, 0, 0.533}
\definecolor{symbolsclr}{rgb}{0.5, 0.25, 0.25}   
\definecolor{commentclr}{rgb}{0, 0.5, 0}       
\definecolor{stringsclr}{rgb}{0.8, 0.4, 0.1}     
\definecolor{numbersclr}{rgb}{0.8, 0.1, 0}
\definecolor{bckgrndclr}{rgb}{0.95, 0.95, 0.96}
\lstdefinelanguage{Jython}%
  { morekeywords={None,and,as,assert,break,class,continue,def,del,elif,%
      else,except,finally,for,from,global,if,import,in,is,lambda,nonlocal,%
      not,or,pass,print,raise,return,try,while,with,yield,repeat},%
    moreprocnamekeys={class,def},%
    classoffset=2,
    alsoletter={0,1,2,3,4,5,6,7,8,9,.,1e,-},
    morekeywords={0,1,2,3,5,7,8,9,10,12,13,15,20,27,200,201,221,350},
    morekeywords={0.0,0.001,0.003,0.005,0.1,0.3,0.4,0.5,0.6,0.7,0.8,0.9999,1.0,10.0},
    morekeywords={1e-8,1e-7,1e-6,-15,-9},
    literate=*{ , }{,}{1},
    classoffset=1,%
    morekeywords={Exception,False,True,abs,all,any,ascii,bin,bool,bytearray,%
      bytes,callable,chr,classmethod,compile,complex,copyright,delattr,dict,%
      dir,divmod,enumerate,eval,exec,exit,float,format,frozenset,getattr,%
      globals,hasattr,hash,help,hex,id,input,int,isinstance,issubclass,iter,%
      len,license,list,locals,map,max,min,next,object,oct,open,ord,pow,%
      property,quit,range,repr,reversed,round,set,setattr,slice,sorted,%
      staticmethod,str,sum,super,tuple,type,vars,zip,inputInt,inputFloat,%
      inputString,msgDlg,askYesNo,enum,clrScr,head,tail,indices,self},%
    classoffset=0,%
    sensitive=true,%
    showstringspaces=false,%
    morecomment=[l]\#,%
    morestring=[b]',%
    morestring=[b]",%
    morestring=[b]{'''},%
    morestring=[b]{"""}%
  }
\newcommand{\floor}[1]{\left\lfloor #1 \right\rfloor} 
\newcommand{\e}{\mathrm{e}}
\newcommand{\eps}{\epsilon}

\makeatletter
\lst@Key{matchrangestart}{f}{\lstKV@SetIf{#1}\lst@ifmatchrangestart}
\def\lst@SkipToFirst{%
    \lst@ifmatchrangestart\c@lstnumber=\numexpr-1+\lst@firstline\fi
    \ifnum \lst@lineno<\lst@firstline
        \def\lst@next{\lst@BeginDropInput\lst@Pmode
        \lst@Let{13}\lst@MSkipToFirst
        \lst@Let{10}\lst@MSkipToFirst}%
        \expandafter\lst@next
    \else
        \expandafter\lst@BOLGobble
    \fi}
\makeatother

\begin{document}
 
\begin{titlepage}
\begin{center}
\includegraphics[width=0.8\textwidth]{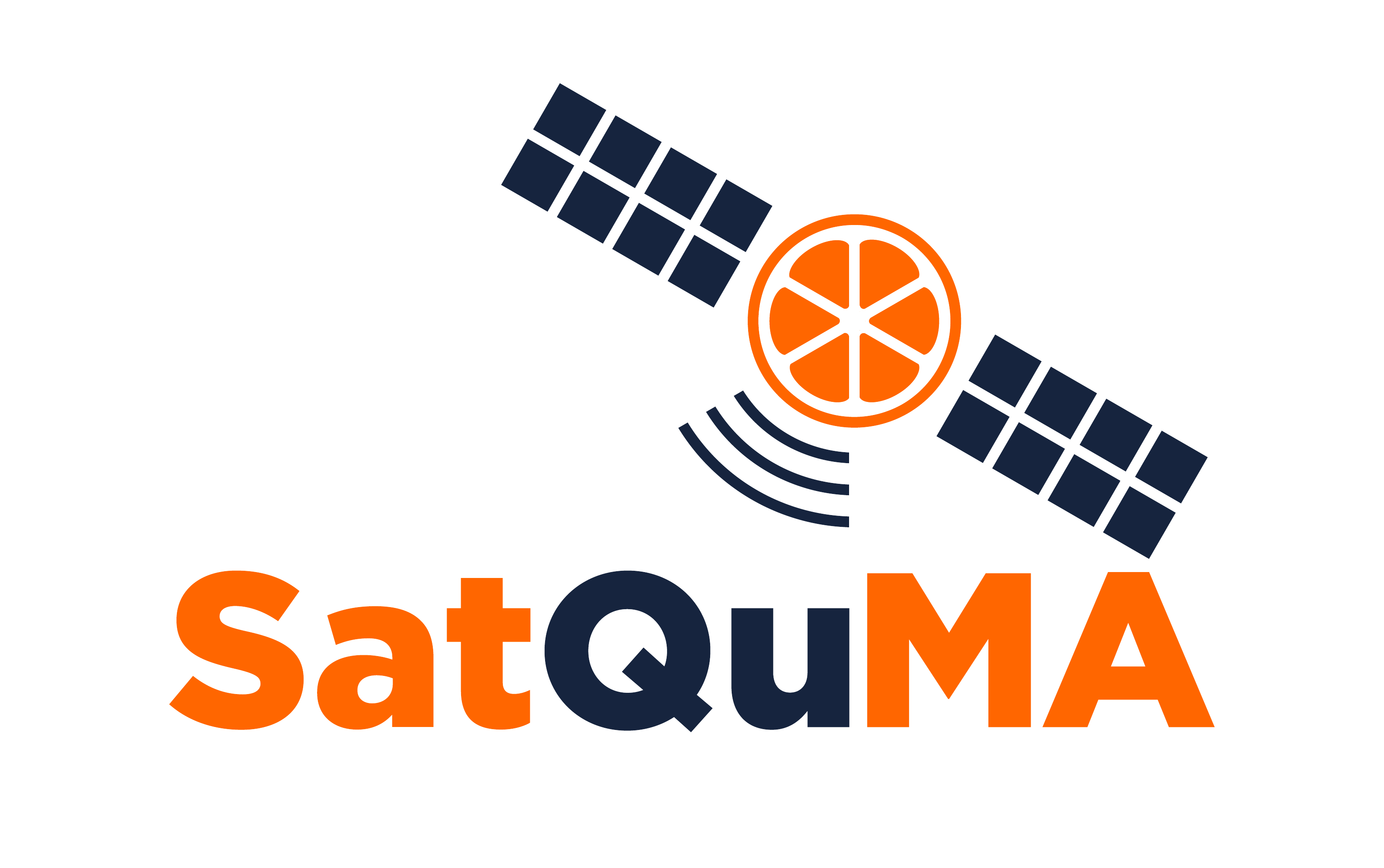}\\[2cm]
{\Large Satellite Quantum Modelling \& Analysis Software}\\
{\large Version 1.1: Documentation}\\[1cm]

{\Large J.~S. Sidhu, T. Brougham, D. McArthur, R.~G. Pousa and D.~K.~L. Oi}\\[1.5cm]

\textsc{\Large University of Strathclyde}\\
\textsc{Department of Physics}\\
\textit{John Anderson Building, 107 Rottenrow East, Glasgow, G4 0NG}\\[1cm]
\vfill
\large December 2021
\end{center}
\end{titlepage}

\chapter{Introduction}
\section{Mission statement}

We provide a numerical key rate analysis, which determines the amount of 
expected key generation in satellite-based quantum key distribution protocols. 
This key length analysis will help develop an intuition on the effects of different 
operational scenarios on the key rate and inform the development of source and 
receiver systems. This numerical toolkit will provide a guide to future 
satellite missions.

\section{Scope of current version}
The current release of SatQuMA, version 1.1, is an update to the first working version that could calculate finite key lengths 
for a limited set of systems and circumstances.
The new version, primarily, provides greatly improved optimisation routines and introduces asymptotic expressions. 
We implement an optimised, asymmetric two-decoy state BB84 protocol with weak coherent pulses.
Elevation, and time, dependent system losses are defined externally and read-in from a data file; an example file is supplied with the software.
We consider multiple satellite overpasses, but only for identical orbits.
Protocol parameters can be either optimised or specified, collectively.

\section{List of updates}
The latest release contains many small improvements over the original v1.0 release, listed here in no particular order:
\begin{itemize}
  \item The system parameters \texttt{Pec} ($p_{ec}$), the probability of extraneous counts, and \texttt{QBERI} ($\text{QBER}_\text{I}$), the intrinsic quantum bit error rate
  (previously called \texttt{Pdc} and \texttt{PolError} respectively) are now iterable by default. 
  \item Due to the introduction of \texttt{Pec} and \texttt{QBERI} as iterable parameters the naming convention for output files has changed such that each file is appended
  with the relevant values of these parameters. Users are now asked to provide the base name (or prefix) for all output files, if requested.
  \item We have introduced a parameter, \texttt{shift\_elev0}, which allows the user to shift the centre of the transmission window from zenith by a specified number of degrees.
  \item We have removed the boolean flag \texttt{tChernoff}, which switched between the Chernoff and Hoeffding tail bounds, and introduced a list of bounds that can be selected
  from - including \texttt{`Chernoff'}, \texttt{`Hoeffding'}, and \texttt{`Asymptotic'} (\textit{i.e.} none). Selecting the \texttt{`Asymptotic'} option for the tail bounds also
  results in some other default parameters being automatically selected, regardless of the users choices, so as to ensure that the asymptotic key rate is calculated properly.  
  \item We have incorporated the other two available constrained optimisation methods from the SciPy package, namely `COBYLA' (Constrained Optimisation BY Linear Approximation)
  and `SLSQP' (Sequential Least SQuares Programming). The former provides a much more robust optimisation than the previous algorithm (`trust-constr') and has been set as the default.
  \item Optimisations for a particular set of parameters are now performed in a while loop, primarily allowing for a brute force approach to finding non-zero keys in high loss/QBER systems.
  This can also greatly improve the performance of the optimiser but at the cost of additional computations. The user can now specify the minimum number of optimisations to be performed
  via the parameter \texttt{NoptMin}. Additionally, there are two boolean flags which can be used to try and minimise the number of optimisations performed. First, the flag \texttt{tStopZero}
  will break out of the while loop if the first \texttt{NoptMin} optimisations all return a value of zero -- this stops the software from wasting time looking for key in parameter regimes where
  there is none. Second, the flag \texttt{tStopBetter} simply breaks out of the while loop if a larger amount of key has been found and the minimum number of optimisations has been satisfied.
  \item The way in which the initial optimisation parameters are set has also been changed. If a user has requested that these parameters are initialised from a set of specified values, then
  these values are used for each calculation -- preciously they were only used for the first calculation. If a user has instead requested that the parameters are to be initialised randomly then, 
  if it is not the first calculation, and the previous calculation did not produce zero key, the parameters from the previous calculation are tried first before being randomised. 
  If it is the first calculation, or the previous calculation did return zero key, then a new set of random parameters are generated. 
  \item We have removed the boolean flag \texttt{tSortData}, which enabled sorting of the data to be written out, as it is not necessary without parallel computations. We have also
  introduced the boolean flag \texttt{tMetrics} which allows the user to explicitly specify whether the optimisation metrics for each (final) calculation should be written to file.
  \item We have introduced two calculation timers; one for the total calculation runtime and one for each block of losses and transmission time windows.
  \item We have corrected a minor bug in the method of determining the size of the data arrays. This bug was only an issue when using an even-valued step size for either the 
  losses (\texttt{ls}) or transmission time half-windows (\texttt{dt}) arrays with an odd-valued end point.
\end{itemize}

\section{Installation}
The latest version of the SatQuMA software can be found at \url{https://github.com/cnqo-qcomms/SatQuMA}.

\subsection{Required packages}
The current version of SatQuMA requires an installation of Python 3.* and the following standard packages:
\begin{itemize}
  \item scipy (SciPy)
  \item numpy (NumPy)
  \item sys
  \item time
\end{itemize}

\chapter{Theoretical summary}

\section{Background \label{sec:theory}}

Here we present a high-level summary of the equations required to calculate the secret key length (SKL) which appear in the current SatQuMA release. 
We do not, however, provide any form of derivation for these relations here.

\subsection{Protocol and statistics}

In our protocol, Alice randomly prepares a state in the basis $X$ or $Z$, where we usually assume $X=\{D,A\}$ and $Z=\{H,V\}$, with one of three intensities $\mu=\{\mu_1,\mu_2,\mu_3\}$,
each with a probability of being selected $P_{\mu}=\{P_{\mu_1},P_{\mu_2},P_{\mu_3}\}$ .

Once Alice has sent her signals to Bob and the reconciliation process, error correction, and post-processing has been completed we can define some measurement statistics from the sifted key.
We define the number of events, for each basis, for each intensity Alice could prepare
\begin{align}
  n_{\text{X},\mu} &:= \{n_{\text{X},\mu_1}, n_{\text{X},\mu_2}, n_{\text{X},\mu_3}\}, \label{eqn:nXmu} \\
  n_{\text{Z},\mu} &:= \{n_{\text{Z},\mu_1}, n_{\text{Z},\mu_2}, n_{\text{Z},\mu_3}\}, \label{eqn:nZmu}
\end{align}
and similarly we define the number of bit errors, for each basis, for each intensity
\begin{align}
  m_{\text{X},\mu} &:= \{m_{\text{X},\mu_1}, m_{\text{X},\mu_2}, m_{\text{X},\mu_3}\}, \label{eqn:mXmu} \\
  m_{\text{Z},\mu} &:= \{m_{\text{Z},\mu_1}, m_{\text{Z},\mu_2}, m_{\text{Z},\mu_3}\}. \label{eqn:mZmu}
\end{align}

\subsection{Secure key length}
The length of the secure key is given by~\cite{Lim_PRA:2014}
\begin{equation}
  \ell = \floor{ s_{\text{X},0} + s_{\text{X},1}\left[1 - h(\phi_\text{X})\right] - \lambda_\text{EC} - 6\log_2\left(\frac{21}{\eps_\text{s}}\right) - \log_{2}\left(\frac{2}{\eps_\text{c}}\right)}, \label{eqn:SKL}
\end{equation}
where the outer brackets indicate that we should take the floor of this expression. Here, $s_{\text{X},0}$ is the number of vacuum events, $s_{\text{X},1}$ is the number of single-photon events, and $\phi_\text{X}$ is the phase error rate in the sifted $X$ basis. The parameter 
$\lambda_\text{EC}$ provides an estimate, or a bound, on the number of bits required for error correction although this should be replaced with the \emph{actual} number of bits used when this is known. The security parameters $\eps_\text{c}$ and $\eps_\text{s}$ are the prescribed security parameters which define the \emph{correctness} and \emph{secrecy} of the resulting key respectively.
The binary entropy function used above is defined as
\begin{equation}
  h\left(x\right) = -x \log_2 x - \left(1 - x\right) \log_2\left(1 - x\right). \label{eqn:h}
\end{equation}

\subsection{Number of vacuum events}

The number of vacuum events in a particular basis is evaluated as, for example,
\begin{equation}
  s_{\text{X},0} \geq \tau_0 \frac{\mu_2 n_{\text{X},\mu_3}^{-} - \mu_3 n_{\text{X},\mu_2}^{+}}{\mu_2 - \mu_3}, \label{eqn:sX0}
\end{equation}
where the probability that Alice sends an $n$-photon state is given by the Poisson distribution,
\begin{equation}
  \tau_n = \sum\limits_{j=1}^{3} \frac{\e^{-\mu_j} \mu_{j}^n p_{j}}{n!}. \label{eqn:tau}
\end{equation}
In order to account for statistical fluctuations in the expected number of $n$-photon events, we apply the Chernoff bound and define the functions~\cite{Yin_SciRep:2020}
\begin{subequations}
\begin{align}
  n_{\text{X},\mu_j}^{+} &= \frac{\e^{\mu_j}}{p_{\mu_j}}\left[n_{\text{X},\mu_j} + \log_{\e}\left(\frac{21}{\eps_\text{s}}\right) + \sqrt{2n_{\text{X},\mu_j} \log_{\e}\left(\frac{21}{\eps_\text{s}}\right) +  \log_{\e}\left(\frac{21}{\eps_\text{s}}\right)^2}  \right],  \label{eqn:nXp} \\
  n_{\text{X},\mu_j}^{-} &= \frac{\e^{\mu_j}}{p_{\mu_j}}\left[n_{\text{X},\mu_j} - \frac{1}{2}\log_{\e}\left(\frac{21}{\eps_\text{s}}\right) - \sqrt{2n_{\text{X},\mu_j} \log_{\e}\left(\frac{21}{\eps_\text{s}}\right) +  \frac{1}{4}\log_{\e}\left(\frac{21}{\eps_\text{s}}\right)^2}  \right].  \label{eqn:nXm}
\end{align}
\end{subequations}
Note, we could also use these expressions to determine the number of vacuum events in the $Z$ basis by exchanging the $n_{\text{X},\mu_j}$ terms for the corresponding $n_{\text{Z},\mu_j}$ terms.

\subsection{Number of single-photon events \label{ssec:s1}}

The number of single-photon events in a particular basis is similarly evaluated as,
\begin{equation}
  s_{\text{X},1} \geq \tau_1 \frac{\mu_1\left[ n_{\text{X},\mu_2}^{-} -  n_{\text{X},\mu_3}^{+} - \frac{\mu_2^2 - \mu_3^2}{\mu_1^2}\left( n_{\text{X},\mu_1}^{+} - \frac{s_{\text{X},0}}{\tau_0}\right)\right]}{\mu_1\left(\mu_2-\mu_3\right) - \mu_2^2 + \mu_3^2}, \label{eqn:sX1}
\end{equation}
where $\tau_1$ is given by (\ref{eqn:tau}). As with the expressions for the number of vacuum events, we can use the same expression to determine the number of single-photon events in the $Z$ basis by again exchanging the $n_{\text{X},\mu_j}$ terms for the corresponding $n_{\text{Z},\mu_j}$ terms.

\subsection{The phase error rate}

We evaluate the phase error rate in the $X$ basis according to
\begin{equation}
  \phi_\text{X} \leq \frac{v_{\text{Z},1}}{s_{\text{Z},1}} + \gamma\left(\eps_\text{s}, \frac{v_{\text{Z},1}}{s_{\text{Z},1}}, s_{\text{Z},1}, s_{\text{X},1}\right), \label{eqn:phiX}
\end{equation} 
where we use the function
\begin{equation}
  \gamma\left(a,b,c,d\right) = \sqrt{\frac{\left(c+d\right)\left(1-b\right)b}{cd\log_\e 2}\log_2\left[\frac{c+d}{bcd\left(1-b\right)} \frac{21^2}{a^2}\right]}, \label{eqn:gamma}
\end{equation}
and the single-photon events are defined as above in Sec.~\ref{ssec:s1}.
We have also introduced the number of bit errors associated with single-photon events in $Z$
\begin{equation}
  v_{\text{Z},1} \leq \tau_1 \frac{ m_{\text{Z},\mu_2}^{+} -  m_{\text{Z},\mu_3}^{-}}{\mu_2 - \mu_3}, \label{eqn:vZ1}
\end{equation}
where the bounds on the number of bit errors due to statistical fluctuations, $m_{\text{Z},\mu_j}^{\pm}$ are given by (\ref{eqn:nXp}) and (\ref{eqn:nXm}) where the number(s) of events in the $X$ basis, $n_{\text{X},\mu_j}$, should be substituted with the number(s) of bit errors in the $Z$ basis, $m_{\text{Z},\mu_j}$.

\subsection{Estimating the amount of error correction}

We estimate the number of bits that need to be sacrificed to perform the error correction in two ways: the first is more accurate but also much more complex; the second is simple to implement but provides only a lower bound.

\subsubsection{Method 1}

We can evaluate the number of bits that we need to sacrifice for error correction as~\cite{Tomamichel_QIP:2017}
\begin{align}
  \lambda_\text{EC} &\approx n_{X} h\left(\text{QBER}_\text{X}\right) + \left[n_{X}\left(1 - \text{QBER}_\text{X}\right) - F^{-1}\left(\eps_\text{c};\floor{n_\text{X}},1-\text{QBER}_\text{X}\right) - 1\right]
\nonumber \\ & \times\log_{\e}\left[\frac{\left(1-\text{QBER}_\text{X}\right)}{\text{QBER}_\text{X}}\right] - \frac{1}{2} \log_{\e} n_\text{X} - \log_{\e} \left(\frac{1}{\eps_\text{c}}\right), \label{eqn:lambdaEC_1}
\end{align}
where we define the quantum bit error rate in the $X$ basis as
\begin{equation}
  \text{QBER}_\text{X} = \frac{\sum_j m_{\text{X},\mu_j}}{\sum_j n_{\text{X},\mu_j}},\; \text{for}\; j\in\{1,2,3\}, \label{eqn:QBERx}
\end{equation}
and $F^{-1}(k;n,p)$ is the inverse (or quantile function) of the binomial cumulative distribution function
\begin{equation*}
  F(k;n,p)=\sum\limits_{i=0}^{\floor{k}}\left(\begin{matrix}n\\i\end{matrix}\right) p^i \left(1-p\right)^{n-i}. 
\end{equation*}

\subsubsection{Method 2}

Another method, based upon the block size, estimates the lower bound on the error correction as
\begin{equation}
  \lambda_\text{EC} \geq 1.16 \sum\limits_{j=1}^{3} n_{\text{X},\mu_j} h\left(\text{QBER}_\text{X}\right). \label{eqn:lambdaEC_2}
\end{equation}

\subsubsection{Method 3}

We can simply estimate the lower bound on the error correction based upon the total number of bit errors in the $X$ basis
\begin{equation}
  \lambda_{EC} \geq 1.16 \sum\limits_{j=1}^{3} m_{X,\mu_j}. \label{eqn:lambdaEC_3}
\end{equation}



\chapter{Satellite overpass geometry}

In this chapter we will discuss the way in which we define a satellite overpass, and how SatQuMA expects a transmission window to be specified for the purposes of a key length calculation.

\begin{figure}
  \centering
  \includegraphics[width=\textwidth]{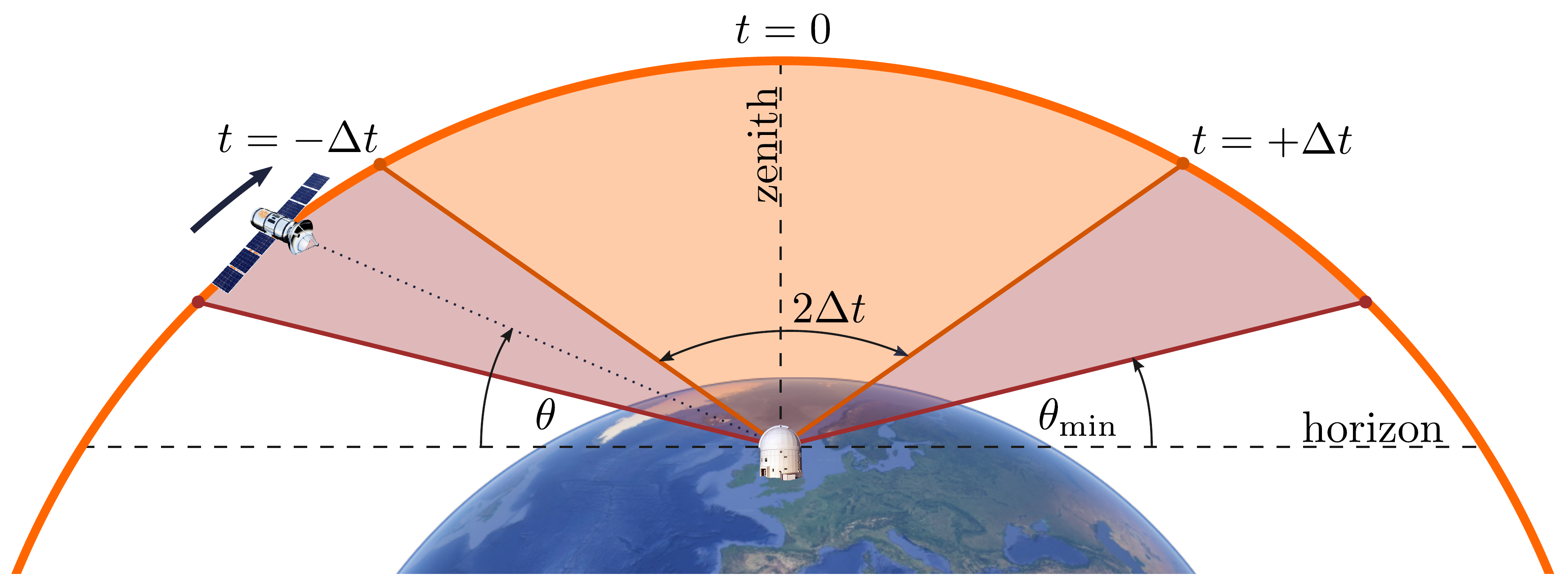}
  \caption{Satellite overpass geometry. A satellite passes over an OGS, where the satellite elevation angle $\theta\in\left[0^\circ,90^\circ\right]$ is measured from the local horizon of the OGS. The satellite and OGS can only close a link when the satellite is above the minimum elevation angle $\theta_\text{min}$. We assume that the actual transmission window is symmetric about the local zenith, which we label as $t=0$, and that signals are sent while $-\Delta t \leq t \leq \Delta t$.\smallskip\\
\tiny{OGS photo: ESA\\
Globe: Google, Data SIO, NOAA, U.S. Navy, NGA, GEBCO, Landsat/Copernicus IBCAO U.S. Geological Survey} \label{fig:sat}}
\end{figure}

In SatQuMA, the transmission window is assumed to be symmetric in time (and elevation) about the local zenith by default and discretized into time-slots. An illustration of an ideal, zenith satellite overpass is shown in Fig.~\ref{fig:sat}. To define the overpass geometry we set a minimum elevation angle, $\theta_\text{min}$, for which signals can be transmitted between the satellite and OGS/receiver and choose a transmission window half-width, $\Delta t$. The software will attempt to generate secret key over each time-slot while $-\Delta t \leq t \leq \Delta t$.

\begin{figure}
  \centering
  \includegraphics[width=0.6\textwidth]{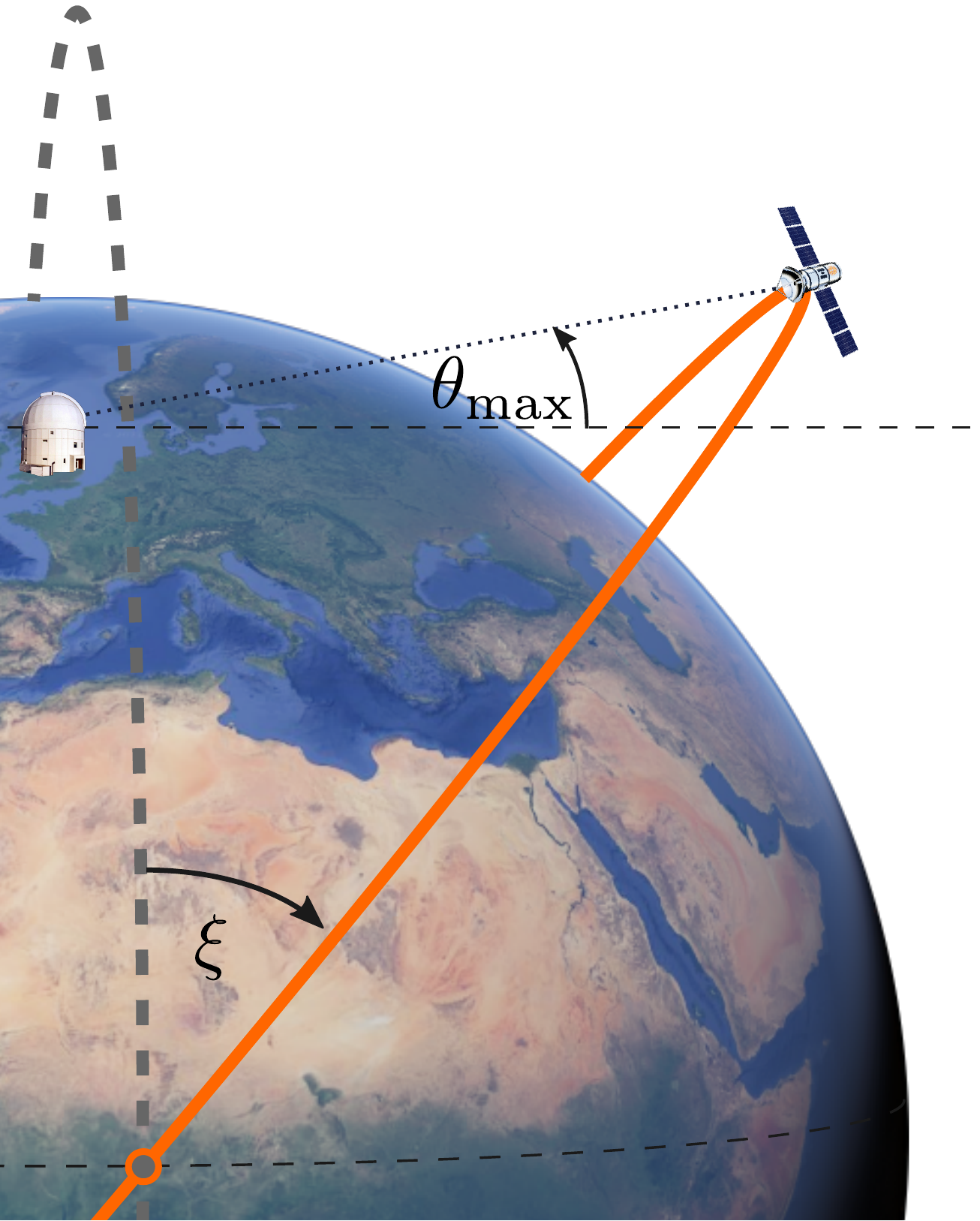}
  \caption{Non-zenith satellite orbits. The satellite orbit can be defined in terms of the maximum elevation that the satellite will reach, $\theta_\text{max}\in\left[0^\circ,90^\circ\right]$, above the local horizon and also the angle with respect to the centre of the Earth, $\xi$, that an ideal zenith orbit needs to be rotated to have equivalent geometry.
\smallskip\\
\tiny{OGS photo: ESA\\
Globe: Google, Data SIO, NOAA, U.S. Navy, NGA, GEBCO, Landsat/Copernicus IBCAO U.S. Geological Survey} 
\label{fig:nonzen}}
\end{figure}

We can also consider non-ideal satellite overpasses, where the orbital geometry can be defined in terms of either the maximum elevation of that orbit, relative to the local horizon of the receiver, or the orbit rotation angle $\xi$, relative to the centre of the Earth, required to transform an ideal zenith overpass into the non-ideal overpass specified.
The relation between these two angles is illustrated in Fig.~\ref{fig:nonzen}.
The maximum elevation is related to the zenith orbit rotation angle $\xi$ as
\begin{equation}
  \theta_\text{max} = \cos^{-1}\left\{\frac{\left(R_\text{E} + h_\text{sat}\right)\sin\xi}
  {\left[\begin{array}{c}
     \left(R_\text{E} + h_\text{sat}\right)^2 + \left(R_\text{E} + h_\text{OGS}\right)^2 - \\
      2\left(R_\text{E} + h_\text{sat}\right)\left(R_\text{E} + h_\text{OGS}\right)\cos\xi\end{array}\right]}\right\}, \label{eqn:tm_xi}
\end{equation}
where $R_\text{E}$ is the radius of the Earth, $h_\text{sat}$ is the orbital altitude of the satellite, and $h_\text{OGS}$ is the altitude of the receiver/ground station.

\chapter{Example of use}
\lstset{numbers=left,matchrangestart=f}
Here we go through the process of setting up an optimisation calculation using SatQuMA, with script excerpts taken from {\ttfamily SatQuMA\_1.1.py} as indicated by the specified line numbers.

Many of the following parameter flags are set by switching between indices of a simple boolean array.
\begin{lstlisting}[firstnumber=54]
F_or_T = [False, True] # List used to switch between False or True values
\end{lstlisting}

\section{Optimisation parameters}
SatQuMA allows the main protocol parameters to be either optimised or specified, collectively.
To enable the optimisation of the protocol parameters we must set the relevant boolean flag to \texttt{True}.
\begin{lstlisting}[firstnumber=72]
tOptimise = F_or_T[1]  # False (0) or True (1)
\end{lstlisting}
The optimiser requires these parameters to be given a range, with an upper and lower bound, which we define using a  {\ttfamily numpy} array.
\begin{lstlisting}[firstnumber=83]
xb = np.array([[0.3,1.0],[0.6,0.9999],[0.0,0.4],[0.3,1.0],[0.1,0.5]])
\end{lstlisting}
Each pair of numbers (columns) of the {\ttfamily numpy} array are the (non-inclusive) lower and upper bounds for the parameters in order $P_x, P_{\mu_1}, P_{\mu_2}, \mu_1, \mu_2$.

We must now set initial values for the optimised parameters, keeping within the pre-defined, respective parameter ranges. 
These initial values can either be directly specified or they can be selected for us randomly. 
We note that initialising the parameters can be an important step as the returned secret key length (SKL) can be zero over relatively broad parameter regions, or indeed in smaller localised regions, which may cause the optimiser to report spurious issues with the supplied function (that it's derivatives appear to be zero).
In order to specify the initial values, we first set the relevant boolean flag to be \texttt{True}.
\begin{lstlisting}[firstnumber=93]
tInit = F_or_T[1]  # False (0) or True (1)
\end{lstlisting}
Next, we set the values for each parameter.
\begin{lstlisting}[firstnumber=99]
Px_i  = 0.5  # Asymmetric polarisation probability
pk1_i = 0.7  # Probability Alice prepares intensity 1
pk2_i = 0.1  # Probability Alice prepares intensity 2
mu1_i = 0.8  # Intensity 1
mu2_i = 0.3  # Intensity 2
\end{lstlisting}
If we instead want the initial parameters to be set randomly we set the following boolean flag to \texttt{False}.
\begin{lstlisting}[firstnumber=93]
tInit = F_or_T[0]  # False (0) or True (1)
\end{lstlisting}

\section{Calculation parameters}

\subsection{Input file options\label{sec:loss_file}}
SatQuMA requires a time/elevation \textit{vs} link efficiency data file (in CSV format) to be specified, hereafter referred to as the `loss file'. 
An example file has been supplied with the software, the name of which we must specify.
\begin{lstlisting}[firstnumber=132]
loss_file = 'FS_loss_XI0.csv'
\end{lstlisting}
We can also specify the directory containing the loss file, if it is not located in the current working directory.
If the loss file is in the current work directory, then we can set this path to be an empty string.
\begin{lstlisting}[firstnumber=130]
loss_path = ''
\end{lstlisting}
Additionally, we can specify the column within the loss file that contains the link efficiency (losses), however the default is column 3.
\begin{lstlisting}[firstnumber=133]
lc = 3 # Column containing loss data in file (counting from 1)
\end{lstlisting}

\subsection{System parameters}
We now specify the relevant parameters which characterise the performance of the system we wish to model.
First, we specify the orbit offset angle $\xi$ (rad.) which define the smallest angle between the orbit plane of the satellite and the zenith plane of the optical ground station (OGS).
\begin{lstlisting}[firstnumber=139]
xi  = 0.0 # Angle between OGS zenith and satellite (from Earth's centre) [rad.]
\end{lstlisting}
This value should be the same as that used when producing the input loss file, as such we typically include the value of $\xi$ in the name of this file.
At present, however, this value is simply included in the output data and doesn't factor into the calculations directly.

Next, we specify the remaining protocol parameters: the intensity of the third weak coherent pulse (second decoy state) $\mu_3$ and the prescribed errors in both correctness and secrecy, $\epsilon_\text{c}$ and $\epsilon_\text{s}$ respectively.
\begin{lstlisting}[firstnumber=141]
mu3   = 0         # Intensity of pulse 3 (fixed)
\end{lstlisting}
\begin{lstlisting}[firstnumber=143]
eps_c = 10**(-15) # Correctness parameter
eps_s = 10**(-9)  # Secrecy parameter
\end{lstlisting}
Here, by setting $\mu_3=0$ we have chosen the third pulse in our protocol to be the vacuum state.

The intrinsic Quantum Bit Error Rate ($\text{QBER}_\text{I}$) is an iterable parameter, and the values to calculate should be passed as an iterable object: list, array, tuple/singleton, or generator
\begin{lstlisting}[firstnumber=146]
QBERI_list = [0.001,0.003,0.005] # list, array, tuple or singleton
\end{lstlisting}
The extraneous count probability ($P_\text{ec}$) is also iterable.
\begin{lstlisting}[firstnumber=148]
Pec_list   = [1e-8,1e-7,1e-6] # list, array, tuple or singleton
\end{lstlisting}
The after-pulse probability of each detector must also be specified.
\begin{lstlisting}[firstnumber=150]
Pap = 0.001       # After-pulse probability
\end{lstlisting}

Finally, we specify the number of (identical) satellite overpasses to include and the repetition rate of the transmission source (Hz).
\begin{lstlisting}[firstnumber=152]
NoPass = 1        # Number of satellite passes
\end{lstlisting}
\begin{lstlisting}[firstnumber=154]
Rrate  = 1*10**(9) # Source rate (Hz)
\end{lstlisting}

\subsection{Time window and system loss\label{sec:dt_SysLoss}}
The first version of SatQuMA has been designed to calculate (loop) over the duration of the overpass time half-window (s) and the relative system loss (dB).
The time window loop is principally controlled by a \texttt{numpy} array specifying the start, stop and step indices relating to the time slots (as specified in the input loss file, see \ref{sec:loss_file}).
\begin{lstlisting}[firstnumber=162]
dt_range = np.array([200, 350, 10]) # Start, stop, step
\end{lstlisting}
Note, these time slots are labelled relative to the overpass zenith for which we arbitrarily set $t=0$.
Here we have requested SKL calculations with overpass transmit time half-windows of, initially, up to 200 s then rising to half-windows of a duration of 350 s in 10 s increments.
We can also set a minimum elevation for transmission (in degrees) which will override the values we have just specified where necessary.
\begin{lstlisting}[firstnumber=165]
min_elev = 10.0 # Minimum elevation transmission angle (degs)
\end{lstlisting}
Here, we have specified that no transmission is possible for elevations below $10^\circ$.
We can also shift the elevation angle taken as the centre of the pass ($t=0$) in degrees.
\begin{lstlisting}[firstnumber=166]
shift_elev = 0.0 # Shift the elevation angle taken as t = 0 (degs)
\end{lstlisting}

Next, we define the start, stop and step values for a loop over the excess system losses. That is, we can add additional loss to those specified in the loss file (converting from system efficiency).
\begin{lstlisting}[firstnumber=169]
ls_range = np.array([0, 12, 2])   # Start, stop, step value
\end{lstlisting}
Here, we have considered systems which have 0 to 12 dB of excess loss above the losses specified in the input loss file.

\subsection{Output file options}
During a calculation, SatQuMA can write to two different output streams: a local file and the standard output (IDE/terminal/screen/\textit{etc}).
We need to specify each output file that we wish to generate.
To request that the full calculation data for a given set of system loss metrics and time windows is written to file (in CSV format) we first set the relevant boolean flag to be \texttt{True}.
\begin{lstlisting}[firstnumber=186]
tFullData  = F_or_T[1]  # False (0) or True (1)
\end{lstlisting}
If we want only the optimal time window data, we can set the following flag.
\begin{lstlisting}[firstnumber=188]
tOptiData  = F_or_T[1]  # False (0) or True (1)
\end{lstlisting}
Likewise, if we want the optimal time window data for each calculation in a single file
\begin{lstlisting}[firstnumber=190]
tMultiOpt   = F_or_T[1]  # False (0) or True (1)
\end{lstlisting}
We can also have SatQuMA write out the optimiser metrics for each calculation
\begin{lstlisting}[firstnumber=192]
tMetrics    = F_or_T[1]  # False (0) or True (1)
\end{lstlisting}

then choose the path and base filename for the output.
\begin{lstlisting}[firstnumber=195]
outpath    = ''    # Path for output file (empty = current directory)
outbase    = 'out' # Name for output file (minus .csv)
\end{lstlisting}
Finally, we request that SatQuMA print out data to the standard output stream as they are calculated.
\begin{lstlisting}[firstnumber=200]
tPrint     = F_or_T[1]  # False (0) or True (1)
\end{lstlisting}

\subsection{Advanced parameters}
We can also specify some of the advanced parameters; these parameters have been set to default values which should only be changed by more advanced users looking for additional control over the calculations as they may strongly affect the software performance and resulting SKL.
We will specify that the protocol uses the Chernoff bounds, for all tail bounds, by selecting the default option from the list of bound functions.
\begin{lstlisting}[firstnumber=214]
boundOpts = ['Chernoff','Hoeffding','Asymptotic']
boundFunc = boundOpts[0] # Select an option from the list above.
\end{lstlisting}
Next we can select the method used to approximate the number of bits required for error correction (EC). Again, we select the default option from the list of EC functions, which are presented in decreasing order of complexity (\textit{i.e.}\ difficulty to optimise) and accuracy.
\begin{lstlisting}[firstnumber=224]
errcorrOpts = ['logM','block','mXtot','None']
errcorrFunc = errcorrOpts[0] # Select a method from the list above.
\end{lstlisting}
In order to evaluate the effect of an EC function on the optimisation process, we can use a flag to compare the results obtained when performing EC during and after optimisation.
By default this option is turned off.
\begin{lstlisting}[firstnumber=228]
tCompareEC  = F_or_T[0]  # False (0) or True (1)
\end{lstlisting}
We note that when comparing the effect of including the estimate of the number of bits required for error correction both during and after optimisation only the former case is included in the output data.

We can choose the SKL optimisation method from the three constrained optimisation algorithms provided with \texttt{scipy} however we strongly recommend that users select the default option \texttt{`COBYLA'} to ensure robust performance.
\begin{lstlisting}[firstnumber=233]
opt_methods = ['COBYLA','SLSQP','trust-constr']
method      = opt_methods[0] # Select a optimisation method
\end{lstlisting}
Due to the strong influence of the initial parameters on the final optimised output we include an option to set the minimum number of optimisations performed for a given set of parameters.
The default value is 10, which tends to be enough for most parameter ranges, and we would recommend a minimum of 5 for users looking to speed up their calculations.
\begin{lstlisting}[firstnumber=235]
NoptMin  = 10
\end{lstlisting}
The actual number of optimisations performed is also partially determined by the number of function evaluations reported by the optimisation algorithm.
There are some additional flags which will stop the repeat optimisation loops for a given parameter set which are best employed when the user expects that there will be 
regions of zero SKL or where the optimised SKL varies little with the initial parameters.
\begin{lstlisting}[firstnumber=237]
tStopZero   = F_or_T[1] # Stop optimizing if the first NoptMin return SKL = 0?
tStopBetter = F_or_T[1] # Stop after NoptMin optimizations if SKL improved?
\end{lstlisting}
The first flag above allows the user to specify if SatQuMA should exit the optimisation loop if the optimised SKL is consistently returned as zero.
The second flag controls whether the optimisation loop should be exited if the optimised SKL has improved during the loop.

\section{Visualising data\label{sec:visual_data}}
\begin{figure}
  \centering
  \includegraphics[width=\textwidth]{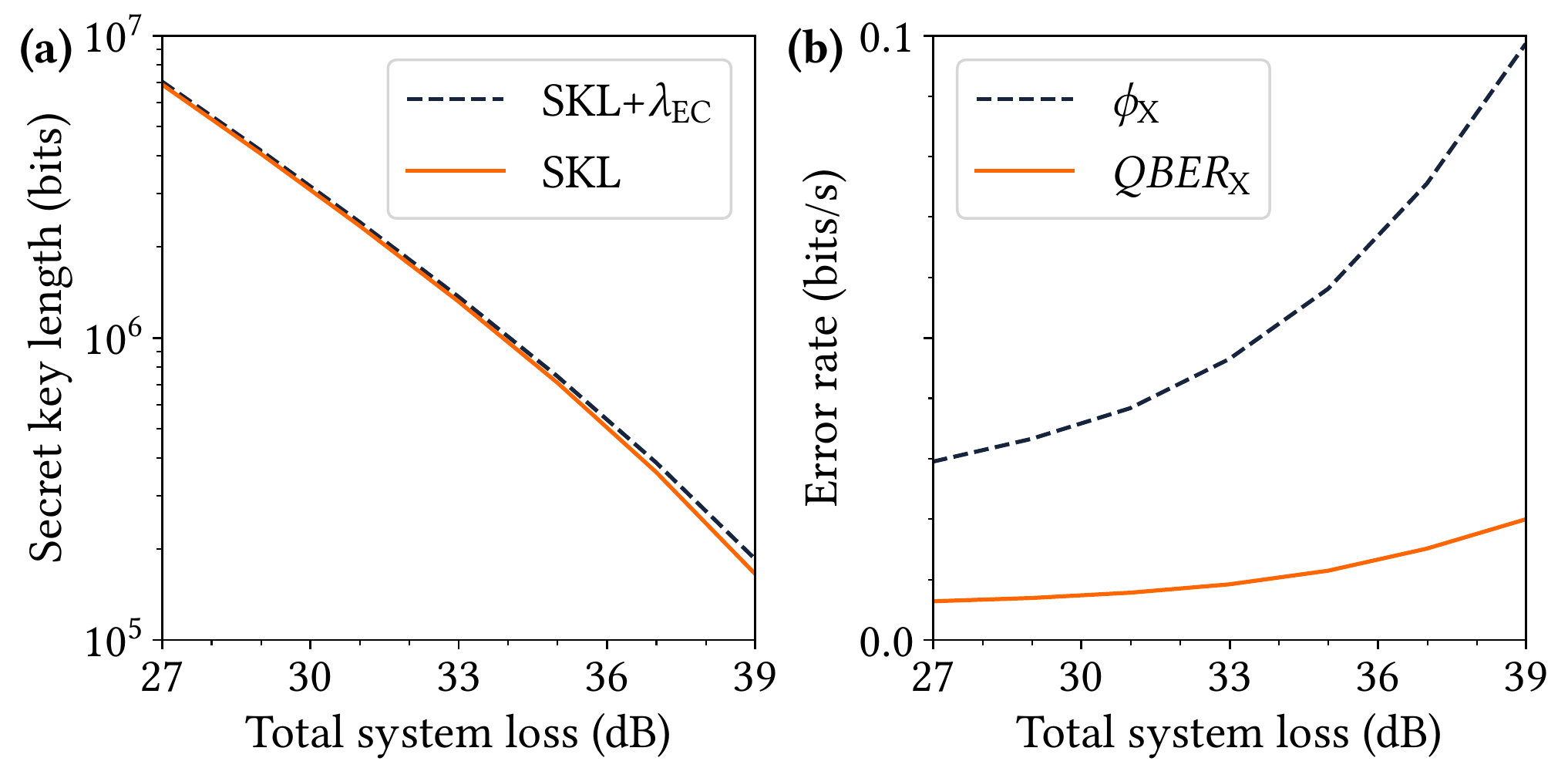}
  \caption{Total system loss in decibels against (a) secret key length, with and without error correction estimation, and (b) the phase and quantum bit error rates for the X basis. \label{fig:ex1}}
\end{figure}

\begin{figure}
  \centering
  \includegraphics[width=0.72\textwidth]{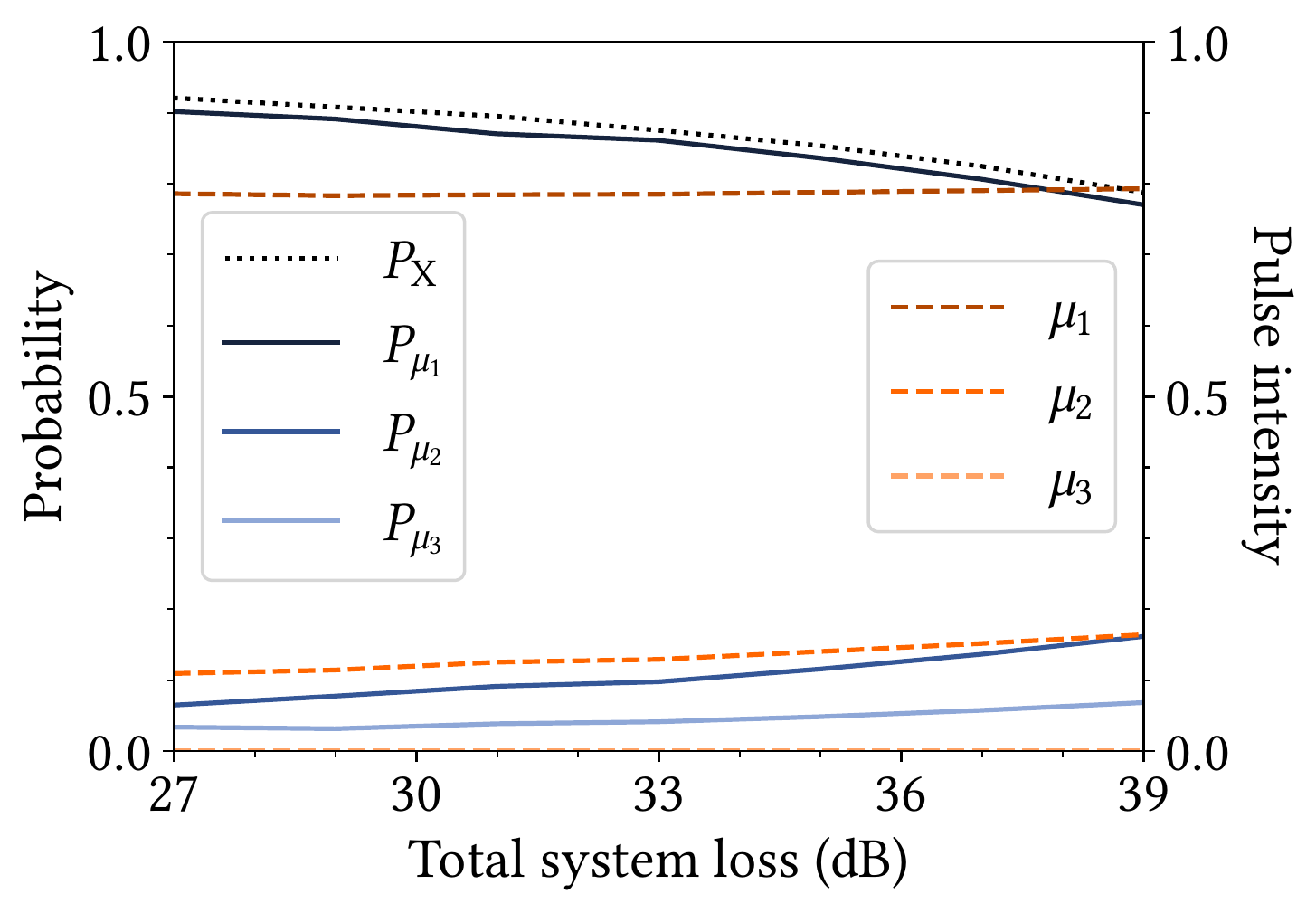}
  \caption{Optimised protocol parameters (for asymmetric BB84) as a function of the total system loss in decibels. Note, $\mu_3=0$. \label{fig:ex2}}
\end{figure}

Once the software has successfully completed the calculation we can visualise the output.
Here we will focus on the optimised, sorted output data file \texttt{out\_opt.csv}.
For example, we can plot the secret key length as a function of the total system loss, both including and excluding the error correction estimation, using the following (minimum working) python code.
\begin{lstlisting}[firstnumber=1]
import numpy as np
data = np.loadtxt('out_opt.csv',skiprows=1,delimiter=',')
x   = data[:,0]      # Total system loss
y1  = data[:,2]      # SKL
y1_ = y1 + data[:,7] # SKL + lambda_EC
import matplotlib.pyplot as plt
fig1, ax1 = plt.subplots(1,1,figsize=(5,5))
ax1.semilogy(x,y1,'-',x,y1_,'--')
\end{lstlisting}
The resulting graph is shown in Fig.~\ref{fig:ex1}(a).
We may also extend our plotting code to visualise the error rates associated with these finite keys, as shown in Fig.~\ref{fig:ex1}(b).
\begin{lstlisting}[firstnumber=9]
fig2, ax2 = plt.subplots(1,1,figsize=(5,5))
y2  = data[:,3:5]    # QBERx and phi_x
ax2.plot(x,y2[:,0],'-',x,y2[:,1],'--')
\end{lstlisting}
Finally, we may also plot the change in the optimised protocol parameters, as shown in Fig.~\ref{fig:ex2}, using the following code.
\begin{lstlisting}[firstnumber=12]
fig3, ax3 = plt.subplots(1,1,figsize=(5,5))
y3  = data[:,20:27]  # Protocol parameters
ls  = [':','-','-','-','--','--','--'] # Linestyles
for ii in range(0,7,1):
    ax3.plot(x,y3[:,ii],ls[ii])
\end{lstlisting}
The various other parameters and output data from this file are listed in Table \ref{tab:data}.

\begin{table}[h]
  \begin{center}
  \begin{tabularx}{\textwidth}{|c|c|c|X|}
    \hline
    Index & Variable & Symbol & Description \\
   \hline
   0 & \texttt{ls+sysLoss} & - &  Total system loss (dB). \\
   1 & \texttt{dt} & $\Delta t$  &  Transmission half-window duration (s). \\
   2 & \texttt{SKL} & $\ell$ &  Secret key length (bits). \\
   3 & \texttt{QBERx} & $\text{QBER}_\text{X}$ & Quantum bit error rate for X basis (bits/s). \\
   4 & \texttt{phi\_x} & $\phi_\text{X}$ & Phase error rate for X basis (bits/s). \\
   5 & \texttt{nX} & $n_\text{X}$ & Number of events in the X basis.  \\
   6 & \texttt{nZ} & $n_\text{Z}$ & Number of events in the Z basis. \\
   7 & \texttt{lambdaEC} & $\lambda_\text{EC}$ & Estimate of number of bits used for error correction. \\
   8 & \texttt{sX0} & $s_{\text{X},0}$ & No.\ of vacuum events for X basis.  \\
   9 & \texttt{sX1} & $s_{\text{X},1}$ &  No.\ of single photon events for X basis. \\
   10 & \texttt{vz1}  & $v_{\text{Z},1}$ & No.\ of bit errors associated with single-photon events in Z basis.  \\
   11 & \texttt{sZ1} & $s_{\text{Z},1}$ & No.\ of single photon events for Z basis. \\
   12 & \texttt{mpn} & $\sum_j \mu_j / 3$ & Mean transmitted photon number. \\
   13 & \texttt{QBERI} & $\text{QBER}_\text{I}$ & Intrinsic quantum bit error rate (\%). \\
   14 & \texttt{Pec} & $P_\text{ec}$  & Extraneous count probability. \\
   15 & \texttt{Pap} & $P_\text{ap}$ & Probability of after-pulse event. \\
   16 & \texttt{NoPass} & $M$ & Number of satellite overpasses. \\
   17 & \texttt{Rrate} & $f_\text{s}$ & Source repetition rate (Hz). \\
   18 & \texttt{eps\_c} & $\epsilon_\text{c}$ & Correctness parameter. \\
   19 & \texttt{eps\_s} & $\epsilon_\text{s}$ & Secrecy parameter. \\
   20 & \texttt{Px} & $P_\text{X}$ & Polarisation bias for X basis. \\
   21 & \texttt{P1} & $P_{\mu_1}$ & Probability of sending pulse 1. \\
   22 & \texttt{P2} & $P_{\mu_2}$  & Probability of sending pulse 2. \\
   23 & \texttt{P3} & $P_{\mu_3}$ & Probability of sending pulse 3. \\
   24 & \texttt{mu1} & $\mu_1$ & Intensity of pulse 1. \\
   25 & \texttt{mu2} & $\mu_2$ & Intensity of pulse 2. \\
   26 & \texttt{mu3} & $\mu_3$ & Intensity of pulse 3. \\
   27 & \texttt{xi} & $\xi$  & Offset angle of satellite orbital plane from OGS zenith (deg). \\
   28 & \texttt{min\_elev} & $\theta_\text{min}$ & Minimum elevation of satellite for transmission (deg).  \\
   29 & \texttt{max\_elev} & $\theta_\text{max}$ & Maximum elevation of satellite overpass (deg). \\
   30 & \texttt{shift\_elev} & $\theta_\text{shift}$ & Angle to shift transmission window central axis (deg). \\
  \hline
  \end{tabularx}
  \caption{Table of data written to main output file(s) giving the index (data column), python variable name, associated mathematical symbol and a brief description. \label{tab:data}}
  \end{center}
\end{table}

\chapter{Bugs and future releases}

SatQuMA is still very much under development with improved versions planned for release in the near future.
The current version employs an out-of-the-box optimisation algorithm, \texttt{trust-constr} from scipy.

\section{Known bugs}
Unfortunately, as the function being minimized does not have derivatives that can be continuously defined for all regions of the relevant parameter space the scipy optimiser can report errors when it encounters locally flat regions of the function space, typically where the SKL drops to zero.

\section{Reporting bugs}
If you encounter any behaviour you consider unexpected, or anytime python raises an exception (excluding for user error), please send a copy of the main SatQuMA python file, and include any relevant output files, to the development team at the University of Strathclyde.
It is always good practice to add python files to an archive (zip/tar/etc) before sending via email to avoid mail being rejected by mail scanning algorithms.

\section{Suggested content}
If there are any features that you would like to see added to this software then please feel free to contact the development team at the University of Strathclyde.
Otherwise, if you are interested in collaborating on a project to either develop or utilise SatQuMA then please contact the development team!

\section{Contact}
Please contact the development team through via Dr D.~K.~L.~Oi at \href{mailto:daniel.oi@strath.ac.uk}{daniel.oi@strath.ac.uk}.


\begin{thebibliography}{1}

\bibitem{Lim_PRA:2014}
C.~C.~W. Lim, M.~Curty, N.~Walenta, F.~Xu, and H.~Zbinden, ``Concise security
  bounds for practical decoy-state quantum key distribution,'' {\em Phys. Rev.
  A}, vol.~89, p.~022307, February 2014.

\bibitem{Yin_SciRep:2020}
H.-L. Yin, M.-G. Zhou, J.~Gu, Y.-M. Xie, Y.-S. Lu, and Z.-B. Chen, ``Tight
  security bounds for decoy-state quantum key distribution,'' {\em Sci. Rep.},
  vol.~10, p.~14312, August 2020.

\bibitem{Tomamichel_QIP:2017}
M.~Tomamichel, J.~Martinez-Mateo, C.~Pacher, and D.~Elkouss, ``Fundamental
  finite key limits for one-way information reconciliation in quantum key
  distribution,'' {\em Quant. Inf. Proc.}, vol.~16, p.~280, October 2017.

\end{thebibliography}
\end{document}